\documentclass[fleqn,usenatbib]{mnras}

\usepackage{newtxtext}

\usepackage[T1]{fontenc}
\usepackage{pdflscape}
\usepackage[figuresright]{rotating}

\DeclareRobustCommand{\VAN}[3]{#2}
\let\VANthebibliography\thebibliography
\def\thebibliography{\DeclareRobustCommand{\VAN}[3]{##3}\VANthebibliography}

\usepackage{graphicx}	
\usepackage{amsmath}	
\usepackage{amssymb}	
\usepackage{pdflscape}

\title[Stealth CME at Mars]{The Impact of a Stealth CME on the Martian Topside Ionosphere}

\author[Smitha Thampi et al.]{
Smitha V. Thampi\thanks{E-mail: smitha\_vt@vssc.gov.in (SVT)},
C. Krishnaprasad, Govind G. Nampoothiri, and Tarun K. Pant
\\
Space Physics Laboratory, Vikram Sarabhai Space Centre, Thiruvananthapuram 695022, 
India
}

\pubyear{2020}

\begin{document}
\maketitle

\begin{abstract}

Solar cycle 24 is one of the weakest solar cycles recorded, but surprisingly the declining phase of 
it  had a  slow CME which evolved without any low coronal signature and is classified as a stealth 
CME which was  responsible for an intense geomagnetic storm at Earth (Dst = -176 nT). The impact of 
this space weather event on the terrestrial ionosphere has been reported. However, the propagation 
of this CME beyond 1 au and the impact of this CME on other planetary environments have not been 
studied so far. In this paper, we analyse the data from Sun-Earth L1 point as well as from the Martian 
orbit (near 1.5 au) to understand the characteristics of the stealth CME as observed beyond 1 au. 
The observations near Earth are using data from the Solar Dynamics Observatory (SDO) and the 
Advanced Composition Explorer (ACE) satellite located at L1 point whereas those near Mars are from 
the instruments for plasma and magnetic field measurements on board Mars Atmosphere and Volatile 
EvolutioN (MAVEN) mission. The observations show that the stealth CME has reached 1.5 au after 7 
days of its initial observations at the Sun and caused depletion in the nightside topside ionosphere 
of Mars, as observed during the inbound phase measurements of the Langmuir Probe and Waves (LPW) 
instrument on board MAVEN. These observations have implications on the ion escape rates from the 
Martian upper atmosphere.

\end{abstract}


\section{Introduction}

Coronal Mass Ejections (CMEs) are eruptions on the Sun, by which solar plasma and magnetic field are 
expelled into the heliosphere. CME eruption processes involve an energy storage phase, which may be 
the product of flux emergence or photospheric flows followed by an energy release phase. There are 
different physical mechanisms proposed for the eruption of CMEs, which include tether cutting or 
flux cancellation mechanism \citep{Moore2001, Amari2003}, shear motion \citep{Aly1990}, kink 
instability \citep{Torok2004}, torus instability \citep{Kliem2006} and magnetic Rayleigh\textendash 
Taylor instability \citep{Mishra2018}. There are different models to explain the solar eruptions 
like flux emergence model \citep{Feynman1995}, Catastrophe model \citep{Forbes1991}, magnetic 
breakout model \citep{Antiochos1999}, reconnection model \citep{Wyper2017} and forced reconnection 
\citep{Srivastava2019}. Based on the morphological evolution, the CMEs are classified as halo CMEs, 
partial halo CMEs, narrow CMEs, and CMEs with low coronal signatures. The CMEs from the Sun, which 
have virtually no identifiable surface or low corona signatures are often referred to as stealth 
CMEs \citep{Robbrecht2009}. These are typically slow CMEs with a speed less than 500 km s$^{-1}$. 
Studies have shown that they can originate either from the quiet Sun region \citep{Ma2010} or from 
an active region \citep{OKane2019}. They can also be originated near the open field lines or coronal 
holes or from faint flux rope eruptions \citep{Adams2014, Lynch2016, Nitta2017}. \citet{Pevtsov2012} 
observed that a plasma channel without a clear filament structure could also become the source 
region of stealth CMEs. A recent study on the magnetic field configuration in which the stealth 
CME occur, show distinct episodes of flare ribbon formation in the stealth CME source active region 
\citep{OKane2020}. In stealth CMEs, the energy storage and release sequence do happen, but the 
energy release is weak, which is probably associated with the magnetic reconnection during the 
eruption or due to an instability process \citep{OKane2019}. The stealth CMEs have no usual solar 
eruption warning signs in the lower corona, making it difficult for space weather predictions and 
therefore these may lead to unpredictable geomagnetic activity and ionospheric storms. 
\citet{Ma2010} have done a statistical analysis of the source location of the CMEs during solar 
minima and reported that almost one third of the CMEs occurring during the solar minimum period are 
of stealth type. \citet{Zhang2007} have studied the connection between the solar eruptions and 
intense magnetic storms on Earth (Dst<-100 nT) during Solar Cycle (SC) 23. They found that 
12\% of the total CMEs were launched without low coronal signatures. When we consider 
the geoeffectiveness of CMEs, several studies have shown that the geoeffectiveness is higher for 
slow CMEs \citep{Ma2010, Lynch2016, Nitta2017}. Since the stealth CMEs are typically of slow 
velocities, understanding their geoeffectiveness is considered to be very important. Typically being 
slow, they spend a long time in the interplanetary space and near{\textendash}space environments  of 
planets, and have high interaction time with other solar wind structures and planetary 
magnetospheres, some of these interactions probably help to enhance their geoeffectiveness 
\citep{Liu2016}. For instance, \citet{Tsurutani2004} found that some slow ICMEs surprisingly caused 
intense geomagnetic storms. However, it is still unclear how slow CMEs lead to enhanced 
geoeffectiveness by interacting with other solar wind structures. Similarly, the impacts of such 
stealth CMEs on environments of planets like Mars are not reported. Since the observations and 
models both show an enhancement in escape rates on unmagnetized planets like Mars and Venus during 
space weather events like CMEs \citep{Jakosky2015b, Brain2016}, understanding the statistics of the 
stealth CMEs and  their impacts are important for quantifying the planetary atmospheric escape 
processes.

The declining phase of the SC\textendash24 had a stealth CME \citep{Mishra2019}, which caused an 
intense geomagnetic storm at Earth (Dst = -176 nT), which is the third most intense storm of the 
SC\textendash24 \citep{Abunin2020, Piersanti2020}. \citet{Astafyeva2020} mentioned it as a 
`{surprise geomagnetic storm}' and studied its impact  on Earth's thermosphere and ionosphere using 
both space\textendash based (the Swarm constellation, GUVI/TIMED) and ground\textendash based (GPS 
receivers, magnetometers, SuperDARN) instruments. However, the arrival and impact of this event on 
other planetary bodies have not been reported yet. In this study, we report the solar wind and 
magnetic field observations from a vantage point near Mars to understand the arrival of this slow 
stealth CME and show the response of Martian topside ionosphere to this event.

\section{Data}

The solar observations are taken from Solar Dynamics Observatory (SDO)/Atmospheric Imaging Assembly 
(AIA;  \citet{Lemen2012}) (\url{https://sdo.gsfc.nasa.gov/}) and the Solar and Heliospheric 
Observatory (SOHO) Large Angle and Spectrometric Coronagraph (LASCO)-C2 (\url 
{https://cdaw.gsfc.nasa.gov/CME_list/}). We have also used the Wang-Sheeley-Arge (WSA)\textendash 
ENLIL+Cone model \citep{Odstrcil2003, Mays2015} from ENLIL Solar Wind Prediction 
(\url{http://helioweather.net/}) for understanding the relative planetary positions and the global 
heliospheric context. The Interplanetary Magnetic Field (IMF) and solar wind speed near 1 au, as 
well as the Sym-H (representing the ring current) variations at Earth are obtained from the NASA 
Space Physics Data Facility (SPDF) OMNIWeb data center (\url{https://omniweb.gsfc.nasa.gov/}). 

The datasets from the Mars Atmosphere and Volatile EvolutioN (MAVEN) instruments are from the 
Planetary Data System (\url{https://pds.nasa.gov/}). The solar wind speed and IMF values near Mars 
are obtained from the Solar Wind Ion Analyzer (SWIA; \citet{Halekas2015}) and 
Magnetometer (MAG; \citet{Connerney2015}) instruments aboard MAVEN spacecraft. SWIA is an energy and 
angular ion spectrometer that measures the energy and angular distributions of solar wind ions of 
energy between 25 eV and 25 keV with 48 logarithmically spaced energy steps. MAG is a fluxgate 
magnetometer that measures the intensity and direction of the IMF. The method to determine the 
upstream solar wind and IMF conditions from MAVEN is described by \citet{Halekas2017}, and is used 
in several studies (e.g. \citet{Lee2017, Krishnaprasad2020}). The Langmuir Probe and Waves (LPW; 
\citet{Andersson2015}) instrument on board MAVEN is used for the in situ electron density and 
electron temperature measurements [Level 2, version 3, revision 01 (V03$\_$R01)]. The Neutral Gas 
and Ion Mass Spectrometer (NGIMS; \citet{Mahaffy2014}) observations of MAVEN are used to understand 
the variations of O$_2^+$ and O$^+$ ion densities in the Martian ionosphere. NGIMS is a quadrupole 
mass spectrometer which measures the composition of neutrals and thermal ions, in the mass range 
2-150 amu with unit mass resolution. The NGIMS Level 2 ion data version 08, revision 01 (V08$\_$R01) 
are used.

\section{Observations}

\subsection{CME event}

Figures 1(a), 1(c), and 1(e) show the images of the solar disk as seen in the 211 \AA{} images from 
the AIA on board SDO on 20 August 2018. These images show the signatures of a filament structure and 
two coronal holes which produces fast solar wind. The quiescent filament structure passed over the 
coronal hole and partially erupted to a Coronal Plasma Channel (CPC). Figure 1(b), 1(d) and 1(f) 
show the region of the coronal plasma channel, at different stages of development.  Several other 
instances of the development of this plasma channel leading to the filament eruption are given in 
\citet{Mishra2019}. It is suggested that the spreading coronal plasma channel might have interacted 
with an open field line of the coronal hole \citep{Mishra2019}, leading to a jet\textendash like 
eruption. The hot coronal plasma channel is visible in other EUV filters of AIA as well 
\citep{Mishra2019}. Following this, a flux rope has also evolved and erupted above the coronal 
plasma channel. So, there are three ejections with very faint evidence in the lower corona, which 
merged with each other to form a complex stealth CME, which traveled through the interplanetary 
space which was observed in the STEREO-A HI-2 (Heliospheric Imager-2) field of view on 24 August 
2018, 08:09 UTC \citep{Mishra2019}. The lower part of the CME interacted with the terrestrial 
magnetosphere on 25 August 2018.  The features of the eruptions, their interplanetary propagation  
and the arrival at Earth are described in detail by \citet{Mishra2019, Abunin2020, Chen2019, 
Piersanti2020}. The HI images (Figure 11, \citet{Mishra2019}) further show that the CME arrived near 
the Mars on 27 August 2018.

Figure 2 shows the WSA-ENLIL+Cone simulation snapshots during the passage of the stealth CME at 
Earth as well as during its arrival at Mars. The color shows the solar wind radial velocity. During 
the CME arrivals at Earth and Mars, the velocity is low. However, there is a high speed stream 
possibly originated from the coronal hole. \citet{Chen2019} reported that after the filament 
eruption, the coronal hole merged with a dimming region on 21 August 2018. This could be the source 
of the fast solar wind stream, which followed the ICME and arrived at the planets.

Figure 3 shows the variation of IMF, solar wind velocity, proton density as well as dynamic 
pressure observed near Mars by MAVEN. The total B as well as the components are shown in Figure 3a. 
For comparison, the Bz values observed at L1 are also shown in the figure. Along with the other 
solar wind parameters observed by MAVEN shown in Figures 3(b-d), the near Earth values are also 
shown for comparison. Apart from this, the Sym-H observed at Earth is also shown to depict the 
occurrence of the intense geomagnetic storm at Earth. It can be seen that, at Earth on 25-26 August 
the IMF Bz shows the signature of a magnetic cloud arriving at Earth. The IMF enhancement at Mars 
starts on 27 August and continues even on 28 August. The peak southward component at Earth is 
$\sim$16 nT, and the total B is as high as 19 nT \citep{Mishra2019}. At Mars, the peak B field 
strength is $\sim$10 nT. When the CME arrived, solar wind velocity near Earth was $\sim$350 km 
s$^{-1}$ which indicated that this was a slowly propagating CME. The solar wind velocity near Earth 
further showed an increase because of the high speed stream, and the peak velocity was observed on 
28 August. Near Mars, the solar wind velocity was $\sim$400 km s$^{-1}$, on the arrival of the ICME. 
The arrival of the high speed stream followed, with peak velocity observed on 28 August. On both 
these planets, by the time the stream arrived, the magnetic field enhancements and fluctuations (due 
to CME) diminished, indicating that the CME already passed. Both at 1 au and 1.5 au, the CME 
structure was therefore bracketed between the ambient slow wind and the high speed stream, thus 
enhancing the effectiveness of interaction. Since MAVEN is in an elliptical orbit, it observes the 
upstream solar wind conditions only intermittently \citep{Halekas2017}, making it difficult to infer 
the exact event arrival time at Mars.  

\subsection{Impact on Martian ionosphere}

During August 2018, the inbound legs of the MAVEN spacecraft were observing the nightside region 
from the near\textendash dusk region to near\textendash midnight sector, and the outbound legs were 
observing the post midnight sector. We make use of MAVEN in-situ observations from the inbound phase 
to understand the response to the ICME. The data from the outbound phase are not used because of the 
low signal levels (characteristic of deep nightside data, due to low plasma concentrations). 
Observations during 24-26 August represent the typical quiet time variation, and the observations on 
27 and 28  August 2018 represent the `event orbits'. 

Table-1 shows the  details of the MAVEN orbits used in this study, such as variation in altitude, 
solar zenith angle (SZA), local solar time (LST), latitude, and longitude during inbound legs. These 
observations pertain to the northern hemisphere of Mars where the influences of the crustal magnetic 
field are a minimum \citep{Acuna1999}. It has been observed that during nightime, the largest 
peak ion densities are found near vertical crustal fields, which form cusps that allow energetic 
electron precipitation, whereas smaller peak densities are found near horizontal crustal fields, 
which hinder energetic electron precipitation into the atmosphere \citep{Girazian2017}. However, 
these effects are observed over the southern hemisphere, where there are strong crustal fields. The 
variability of electron density for the present event are mostly free from these effects, since the 
observations shown here are for the inbound leg, which cover northern hemisphere.

Figure 4a shows several LPW orbits during the event, compared to the quiet time orbits. The quiet 
time orbit data are shown along with the mean and standard deviation (error bars). The third, 
fourth, fifth and sixth orbits on 27 August 2018, and the first orbit on 28 August 2018 show 
significant difference from the quiet time behavior. Above 200 km, the topside electron densities 
are completely depleted during these orbits. There are a few data points in the electron density 
profile around 300-350 km altitude region in the profile corresponding to orbit 6 on 27 August 
2018. However, these are points with very low density values. It is reported that the 
signal-to-noise ratios are reduced below electron densities of $\sim$200 cm$^{-3}$ 
\citep{Fowler2015}. Therefore, we do not infer any information from these isolated structures. At 
150-200 km, we only show that the topside electron densities are completely depleted, compared to 
`quiet orbits' during the space weather event. These gradients are similar to the ionopause-like 
density gradient reported earlier \citep{Vogt2015}. Figure 4b shows the NGIMS observations of 
O$_2^+$ and O$^+$ ion concentrations for the same period. It must be noted that NGIMS alternates 
between ion and neutral modes, whereas LPW measures the electron density in all orbits, and hence 
the signature is seen only in fewer orbits in NGIMS data. Similarly, the number of `quiet' time 
profiles are also fewer for the NGIMS observations, and hence the mean and the standard deviations 
are not given. However, the feature that the topside ion densities are highly depleted during the 
ICME period is unmistakably seen in the NGIMS observations as well. Figure 4c shows the electron 
temperature observations from the LPW measurement during the event period, along with the quiet time 
profiles. The topside electron temperatures are enhanced during all the orbits where electron 
density showed depletion. However, it may be noted that reduced signal-to-noise ratios at regions 
where electron densities are below densities of $\sim$200 cm$^{-3}$ also result in LP temperature 
measurement errors increasing to 100\% or more \citep{Fowler2015}, and therefore we cannot infer 
these as the accurate profiles of T$_e$ during these days. Despite this, it is evident that the 
profiles during the `event orbits' show trends which are significantly different (with enhanced 
values) compared to `quiet orbits'.

\section{Discussion}

The CME event observed near the Sun on 20 August 2018 was a  CME without a preceding shock, and is 
classified as a stealth CME \citep{Mishra2019}. The observations show that while reaching Mars, the 
maximum IMF was $\sim$10 nT, which may be considered as an intense space weather condition at Mars. 
The solar wind velocity observed near Mars was $\sim$400 km s$^{-1}$, and this was a slow CME 
inside a compression region between slow and fast solar winds, even when it reached Mars.

The slow, stealth CME impacted Martian topside ionosphere, and the nightside plasma measurements 
show that the topside thermal ionosphere is significantly depleted. The electron temperature 
measurements showed enhancements during this event period. Similar observations were reported by 
\citet{Cravens1982} for the Venusian nightside ionosphere. On days when disappearing ionospheres 
were observed by the OETP (Orbiter Electron Temperature Probe) aboard Pioneer Venus mission, the 
solar wind dynamic pressure were considerably larger than average. It was shown that depleted  and 
variable plasma densities throughout all or a major part of the nightside Venusian ionosphere 
occurred during periods of large, coherent and horizontal magnetic field events and associated with 
large solar wind dynamic pressures. It was suggested that because  dayside ionopause is at low 
altitudes when the solar wind dynamic pressure is large and the IMF is strong, the  nightside 
ionosphere supplied by the day-to-night transport of plasma disappears. If the  dayside ionosphere 
is severely reduced then it is expected that the supply of ions to the nightside will be  curtailed, 
and the large horizontal magnetic field will inhibit the downward diffusion. As a result of these, 
the night side ionosphere will be disappeared. The present observations show that the same is true 
for Martian ionosphere also.  It may also be noted that since this was a slow CME with bulk solar 
wind velocity near Mars $\sim$400 km s$^{-1}$, the outward flow could be weaker compared to the CMEs 
with larger velocities. Even though the peak dynamic pressure was only $\sim$5 nPa, which is smaller 
compared to the strong CMEs like the March 2015 event \citep{Jakosky2015b, Thampi2018}, the slow 
velocities might have allowed more interaction time, and therefore the effectiveness might have 
increased. This is also due to the fact that the CME was actually within the compression region 
between the fast and slow solar wind.

\section{Summary}

The declining  phase of solar cycle 24 had a slow stealth CME which was responsible for an intense 
geomagnetic storm at Earth with Dst$_{min}$ of -176 nT. The propagation of this CME beyond 1 au and 
the impact of this CME on Martian plasma  environments are studied. The observations show that the 
stealth CME has reached 1.5 au after 7 days of its initial observations at the Sun, with a peak 
magnetic field of $\sim$10 nT . This CME caused depletion in the nightside topside ionosphere of 
Mars. The topside ionosphere also had higher electron temperatures compared to the `quiet' 
values. Even with a peak dynamic pressure as low as 5 nPa, the CME had efficiently impacted the 
Martian ionosphere, because the CME was slow, and was bracketed between the fast and slow solar 
winds. This is an unique example to show how slow CMEs can affect the Martian ionosphere. As almost 
one third of the CMEs occurring during the solar minimum period are of slow, stealth type 
\citep{Ma2010}, characterizing their impact on Martian ionosphere is important for constraining the 
ion escape rates.

\section*{Acknowledgements}

The work is supported by the Indian Space Research Organisation (ISRO).   We thank the staff of the 
ACE Science Center for providing the ACE data and OMNIWeb team for providing the IMF and solar wind 
data.  We gratefully acknowledge the MAVEN team for the data. We also acknowledge using solar 
observations from SDO/AIA. The WSA-ENLIL+Cone model simulations are used from ENLIL Solar Wind 
Prediction. C. Krishnaprasad acknowledges the financial assistance provided by ISRO through a 
research fellowship. This research has made use of SunPy v2.0, an open-source and free 
community-developed solar data analysis Python package (\url{https://sunpy.org/}).

\section*{Data Availability}
The solar wind velocity and IMF at L1 point are obtained from the SPDF OMNIWeb data center 
(\url{https://omniweb.gsfc.nasa.gov/}).The MAVEN data used in this work are taken from the NASA 
Planetary Data System (\url{https://pds.nasa.gov/}). The solar observations are available at  
SDO/AIA (\url{https://sdo.gsfc.nasa.gov/}). The WSA-ENLIL+Cone model simulations are used from 
ENLIL Solar Wind Prediction (\url{https://helioweather.net/}).




\begin{thebibliography}{}
\makeatletter
\relax
\def\mn@urlcharsother{\let\do\@makeother \do\$\do\&\do\#\do\^\do\_\do\%\do\~}
\def\mn@doi{\begingroup\mn@urlcharsother \@ifnextchar [ {\mn@doi@}
  {\mn@doi@[]}}
\def\mn@doi@[#1]#2{\def\@tempa{#1}\ifx\@tempa\@empty \href
  {http://dx.doi.org/#2} {doi:#2}\else \href {http://dx.doi.org/#2} {#1}\fi
  \endgroup}
\def\mn@eprint#1#2{\mn@eprint@#1:#2::\@nil}
\def\mn@eprint@arXiv#1{\href {http://arxiv.org/abs/#1} {{\tt arXiv:#1}}}
\def\mn@eprint@dblp#1{\href {http://dblp.uni-trier.de/rec/bibtex/#1.xml}
  {dblp:#1}}
\def\mn@eprint@#1:#2:#3:#4\@nil{\def\@tempa {#1}\def\@tempb {#2}\def\@tempc
  {#3}\ifx \@tempc \@empty \let \@tempc \@tempb \let \@tempb \@tempa \fi \ifx
  \@tempb \@empty \def\@tempb {arXiv}\fi \@ifundefined
  {mn@eprint@\@tempb}{\@tempb:\@tempc}{\expandafter \expandafter \csname
  mn@eprint@\@tempb\endcsname \expandafter{\@tempc}}}

\bibitem[\protect\citeauthoryear{Abunin, Abunina, Belov  \& Chertok}{Abunin
  et~al.}{2020}]{Abunin2020}
Abunin A.~A.,  Abunina M.~A.,  Belov A.~V.,   Chertok I.~M.,  2020, \mn@doi
  [Solar Physics] {10.1007/s11207-019-1574-8}, 295

\bibitem[\protect\citeauthoryear{Acu{\~n}a et~al.,}{Acu{\~n}a
  et~al.}{1999}]{Acuna1999}
Acu{\~n}a M.~H.,  et~al., 1999, \mn@doi [Science]
  {10.1126/science.284.5415.790}, 284, 790

\bibitem[\protect\citeauthoryear{Adams, Sterling, Moore  \& Gary}{Adams
  et~al.}{2014}]{Adams2014}
Adams M.,  Sterling A.~C.,  Moore R.~L.,   Gary G.~A.,  2014, \mn@doi [The
  Astrophysical Journal] {10.1088/0004-637x/783/1/11}, 783, 11

\bibitem[\protect\citeauthoryear{Aly}{Aly}{1990}]{Aly1990}
Aly J.,  1990, \mn@doi [Computer Physics Communications]
  {https://doi.org/10.1016/0010-4655(90)90152-Q}, 59, 13

\bibitem[\protect\citeauthoryear{Amari, Luciani, Aly, Mikic  \& Linker}{Amari
  et~al.}{2003}]{Amari2003}
Amari T.,  Luciani J.~F.,  Aly J.~J.,  Mikic Z.,   Linker J.,  2003, \mn@doi
  [The Astrophysical Journal] {10.1086/345501}, 585, 1073

\bibitem[\protect\citeauthoryear{Andersson et~al.,}{Andersson
  et~al.}{2015}]{Andersson2015}
Andersson L.,  et~al., 2015, \mn@doi [Space Science Reviews]
  {10.1007/s11214-015-0194-3}, 195, 173

\bibitem[\protect\citeauthoryear{Antiochos, DeVore  \& Klimchuk}{Antiochos
  et~al.}{1999}]{Antiochos1999}
Antiochos S.~K.,  DeVore C.~R.,   Klimchuk J.~A.,  1999, \mn@doi [The
  Astrophysical Journal] {10.1086/306563}, 510, 485

\bibitem[\protect\citeauthoryear{Astafyeva, Bagiya, Förster  \&
  Nishitani}{Astafyeva et~al.}{2020}]{Astafyeva2020}
Astafyeva E.,  Bagiya M.~S.,  Förster M.,   Nishitani N.,  2020, \mn@doi
  [Journal of Geophysical Research: Space Physics]
  {https://doi.org/10.1029/2019JA027261}, 125, e2019JA027261

\bibitem[\protect\citeauthoryear{Brain, Bagenal, Ma, Nilsson  \&
  Stenberg~Wieser}{Brain et~al.}{2016}]{Brain2016}
Brain D.~A.,  Bagenal F.,  Ma Y.-J.,  Nilsson H.,   Stenberg~Wieser G.,  2016,
  \mn@doi [Journal of Geophysical Research: Planets]
  {https://doi.org/10.1002/2016JE005162}, 121, 2364

\bibitem[\protect\citeauthoryear{Chen, Liu, Wang, Zhao, Hu  \& Zhu}{Chen
  et~al.}{2019}]{Chen2019}
Chen C.,  Liu Y.~D.,  Wang R.,  Zhao X.,  Hu H.,   Zhu B.,  2019, \mn@doi [The
  Astrophysical Journal] {10.3847/1538-4357/ab3f36}, 884, 90

\bibitem[\protect\citeauthoryear{Connerney, Espley, Lawton, Murphy, Odom,
  Oliversen  \& Sheppard}{Connerney et~al.}{2015}]{Connerney2015}
Connerney J. E.~P.,  Espley J.,  Lawton P.,  Murphy S.,  Odom J.,  Oliversen
  R.,   Sheppard D.,  2015, \mn@doi [Space Science Reviews]
  {10.1007/s11214-015-0169-4}, 195, 257

\bibitem[\protect\citeauthoryear{Cravens et~al.,}{Cravens
  et~al.}{1982}]{Cravens1982}
Cravens T.,  et~al., 1982, \mn@doi [Icarus]
  {https://doi.org/10.1016/0019-1035(82)90083-5}, 51, 271

\bibitem[\protect\citeauthoryear{Feynman \& Martin}{Feynman \&
  Martin}{1995}]{Feynman1995}
Feynman J.,  Martin S.~F.,  1995, \mn@doi [Journal of Geophysical Research:
  Space Physics] {https://doi.org/10.1029/94JA02591}, 100, 3355

\bibitem[\protect\citeauthoryear{{Forbes} \& {Isenberg}}{{Forbes} \&
  {Isenberg}}{1991}]{Forbes1991}
{Forbes} T.~G.,  {Isenberg} P.~A.,  1991, \mn@doi [\apj] {10.1086/170051},
  \href {https://ui.adsabs.harvard.edu/abs/1991ApJ...373..294F} {373, 294}

\bibitem[\protect\citeauthoryear{Fowler et~al.,}{Fowler
  et~al.}{2015}]{Fowler2015}
Fowler C.~M.,  et~al., 2015, \mn@doi [Geophysical Research Letters]
  {10.1002/2015GL065267}, 42, 8854

\bibitem[\protect\citeauthoryear{Girazian, Mahaffy, Lillis, Benna, Elrod  \&
  Jakosky}{Girazian et~al.}{2017}]{Girazian2017}
Girazian Z.,  Mahaffy P.~R.,  Lillis R.~J.,  Benna M.,  Elrod M.,   Jakosky
  B.~M.,  2017, \mn@doi [Journal of Geophysical Research: Space Physics]
  {10.1002/2016JA023508}, 122, 4712

\bibitem[\protect\citeauthoryear{Halekas et~al.,}{Halekas
  et~al.}{2015}]{Halekas2015}
Halekas J.~S.,  et~al., 2015, \mn@doi [Space Science Reviews]
  {10.1007/s11214-013-0029-z}, 195, 125

\bibitem[\protect\citeauthoryear{Halekas et~al.,}{Halekas
  et~al.}{2017}]{Halekas2017}
Halekas J.~S.,  et~al., 2017, \mn@doi [Journal of Geophysical Research: Space
  Physics] {10.1002/2016JA023167}, 122, 547

\bibitem[\protect\citeauthoryear{Jakosky et~al.,}{Jakosky
  et~al.}{2015}]{Jakosky2015b}
Jakosky B.~M.,  et~al., 2015, \mn@doi [Science] {10.1126/science.aad0210}, 350

\bibitem[\protect\citeauthoryear{Kliem \& T\"or\"ok}{Kliem \&
  T\"or\"ok}{2006}]{Kliem2006}
Kliem B.,  T\"or\"ok T.,  2006, \mn@doi [Phys. Rev. Lett.]
  {10.1103/PhysRevLett.96.255002}, 96, 255002

\bibitem[\protect\citeauthoryear{Krishnaprasad, Thampi, Bhardwaj, Lee, Kumar
  \& Pant}{Krishnaprasad et~al.}{2020}]{Krishnaprasad2020}
Krishnaprasad C.,  Thampi S.~V.,  Bhardwaj A.,  Lee C.~O.,  Kumar K.~K.,   Pant
  T.~K.,  2020, \mn@doi [The Astrophysical Journal] {10.3847/1538-4357/abb137},
  902, 13

\bibitem[\protect\citeauthoryear{Lee et~al.,}{Lee et~al.}{2017}]{Lee2017}
Lee C.~O.,  et~al., 2017, \mn@doi [Journal of Geophysical Research: Space
  Physics] {10.1002/2016JA023495}, 122, 2768

\bibitem[\protect\citeauthoryear{{Lemen} et~al.,}{{Lemen}
  et~al.}{2012}]{Lemen2012}
{Lemen} J.~R.,  et~al., 2012, \mn@doi [Solar Physics]
  {10.1007/s11207-011-9776-8}, \href
  {https://ui.adsabs.harvard.edu/abs/2012SoPh..275...17L} {275, 17}

\bibitem[\protect\citeauthoryear{Liu, Hu, Wang, Luhmann, Richardson, Yang  \&
  Wang}{Liu et~al.}{2016}]{Liu2016}
Liu Y.~D.,  Hu H.,  Wang C.,  Luhmann J.~G.,  Richardson J.~D.,  Yang Z.,
  Wang R.,  2016, \mn@doi [The Astrophysical Journal Supplement Series]
  {10.3847/0067-0049/222/2/23}, 222, 23

\bibitem[\protect\citeauthoryear{Lynch, Masson, Li, DeVore, Luhmann, Antiochos
  \& Fisher}{Lynch et~al.}{2016}]{Lynch2016}
Lynch B.~J.,  Masson S.,  Li Y.,  DeVore C.~R.,  Luhmann J.~G.,  Antiochos
  S.~K.,   Fisher G.~H.,  2016, \mn@doi [Journal of Geophysical Research: Space
  Physics] {https://doi.org/10.1002/2016JA023432}, 121, 10,677

\bibitem[\protect\citeauthoryear{Ma, Attrill, Golub  \& Lin}{Ma
  et~al.}{2010}]{Ma2010}
Ma S.,  Attrill G. D.~R.,  Golub L.,   Lin J.,  2010, \mn@doi [The
  Astrophysical Journal] {10.1088/0004-637x/722/1/289}, 722, 289

\bibitem[\protect\citeauthoryear{Mahaffy, Benna  \& {et al}}{Mahaffy
  et~al.}{2014}]{Mahaffy2014}
Mahaffy P.~R.,  Benna M.,   {et al} 2014, \mn@doi [Space Sci. Rev]
  {10.1007/s11214-014-0091-1}, 185

\bibitem[\protect\citeauthoryear{Mays et~al.,}{Mays et~al.}{2015}]{Mays2015}
Mays M.~L.,  et~al., 2015, \mn@doi [Solar Physics] {10.1007/s11207-015-0692-1},
  290, 1775

\bibitem[\protect\citeauthoryear{Mishra \& Srivastava}{Mishra \&
  Srivastava}{2019}]{Mishra2019}
Mishra S.~K.,  Srivastava A.~K.,  2019, \mn@doi [Solar Physics]
  {10.1007/s11207-019-1560-1}, 294

\bibitem[\protect\citeauthoryear{Mishra, Singh, Kayshap  \& Srivastava}{Mishra
  et~al.}{2018}]{Mishra2018}
Mishra S.~K.,  Singh T.,  Kayshap P.,   Srivastava A.~K.,  2018, \mn@doi [The
  Astrophysical Journal] {10.3847/1538-4357/aaae03}, 856, 86

\bibitem[\protect\citeauthoryear{Moore, Sterling, Hudson  \& Lemen}{Moore
  et~al.}{2001}]{Moore2001}
Moore R.~L.,  Sterling A.~C.,  Hudson H.~S.,   Lemen J.~R.,  2001, \mn@doi [The
  Astrophysical Journal] {10.1086/320559}, 552, 833

\bibitem[\protect\citeauthoryear{Nitta \& Mulligan}{Nitta \&
  Mulligan}{2017}]{Nitta2017}
Nitta N.~V.,  Mulligan T.,  2017, \mn@doi [Solar Physics]
  {10.1007/s11207-017-1147-7}, 292

\bibitem[\protect\citeauthoryear{O'Kane, Green, Long  \& Reid}{O'Kane
  et~al.}{2019}]{OKane2019}
O'Kane J.,  Green L.,  Long D.~M.,   Reid H.,  2019, \mn@doi [The Astrophysical
  Journal] {10.3847/1538-4357/ab371b}, 882, 85

\bibitem[\protect\citeauthoryear{O'Kane, Cormack, Mandrini, Démoulin, Green,
  Long  \& Valori}{O'Kane et~al.}{2020}]{OKane2020}
O'Kane J.,  Cormack C.~M.,  Mandrini C.~H.,  Démoulin P.,  Green L.~M.,  Long
  D.~M.,   Valori G.,  2020, The Magnetic Environment of a Stealth Coronal Mass
  Ejection (\mn@eprint {arXiv} {2012.03757})

\bibitem[\protect\citeauthoryear{Odstrcil}{Odstrcil}{2003}]{Odstrcil2003}
Odstrcil D.,  2003, \mn@doi [Advances in Space Research]
  {10.1016/S0273-1177(03)00332-6}, 32, 497

\bibitem[\protect\citeauthoryear{Pevtsov, Panasenco  \& Martin}{Pevtsov
  et~al.}{2012}]{Pevtsov2012}
Pevtsov A.~A.,  Panasenco O.,   Martin S.~F.,  2012, \mn@doi [Solar Physics]
  {10.1007/s11207-011-9881-8}, 277, 185

\bibitem[\protect\citeauthoryear{Piersanti et~al.,}{Piersanti
  et~al.}{2020}]{Piersanti2020}
Piersanti M.,  et~al., 2020, \mn@doi [Annales Geophysicae]
  {10.5194/angeo-38-703-2020}, 38, 703

\bibitem[\protect\citeauthoryear{Robbrecht, Patsourakos  \&
  Vourlidas}{Robbrecht et~al.}{2009}]{Robbrecht2009}
Robbrecht E.,  Patsourakos S.,   Vourlidas A.,  2009, \mn@doi [The
  Astrophysical Journal] {10.1088/0004-637x/701/1/283}, 701, 283

\bibitem[\protect\citeauthoryear{Srivastava et~al.,}{Srivastava
  et~al.}{2019}]{Srivastava2019}
Srivastava A.~K.,  et~al., 2019, \mn@doi [The Astrophysical Journal]
  {10.3847/1538-4357/ab4a0c}, 887, 137

\bibitem[\protect\citeauthoryear{Thampi, Krishnaprasad, Bhardwaj, Lee,
  Choudhary  \& Pant}{Thampi et~al.}{2018}]{Thampi2018}
Thampi S.~V.,  Krishnaprasad C.,  Bhardwaj A.,  Lee Y.,  Choudhary R.~K.,
  Pant T.~K.,  2018, \mn@doi [Journal of Geophysical Research: Space Physics]
  {10.1029/2018JA025444}, 123, 6917

\bibitem[\protect\citeauthoryear{{T{\"o}r{\"o}k} \& {Kliem}}{{T{\"o}r{\"o}k} \&
  {Kliem}}{2004}]{Torok2004}
{T{\"o}r{\"o}k} T.,  {Kliem} B.,  2004, in {Walsh} R.~W.,  {Ireland} J.,
  {Danesy} D.,   {Fleck} B.,  eds,  ESA Special Publication Vol. 575, SOHO 15
  Coronal Heating. p.~56

\bibitem[\protect\citeauthoryear{Tsurutani, Gonzalez, Zhou, Lepping  \&
  Bothmer}{Tsurutani et~al.}{2004}]{Tsurutani2004}
Tsurutani B.,  Gonzalez W.,  Zhou X.-Y.,  Lepping R.,   Bothmer V.,  2004,
  \mn@doi [Journal of Atmospheric and Solar-Terrestrial Physics]
  {https://doi.org/10.1016/j.jastp.2003.09.007}, 66, 147

\bibitem[\protect\citeauthoryear{Vogt et~al.,}{Vogt et~al.}{2015}]{Vogt2015}
Vogt M.~F.,  et~al., 2015, \mn@doi [Geophysical Research Letters]
  {10.1002/2015GL065269}, 42, 8885

\bibitem[\protect\citeauthoryear{Wyper, Antiochos  \& DeVore}{Wyper
  et~al.}{2017}]{Wyper2017}
Wyper P.~F.,  Antiochos S.~K.,   DeVore C.~R.,  2017, \mn@doi [Nature]
  {10.1038/nature22050}, 544, 452

\bibitem[\protect\citeauthoryear{Zhang et~al.,}{Zhang et~al.}{2007}]{Zhang2007}
Zhang J.,  et~al., 2007, \mn@doi [Journal of Geophysical Research: Space
  Physics] {https://doi.org/10.1029/2007JA012321}, 112

\makeatother
\end{thebibliography}


\begin{landscape}
\begin{figure}
\centering
 \includegraphics[width=0.65\columnwidth]{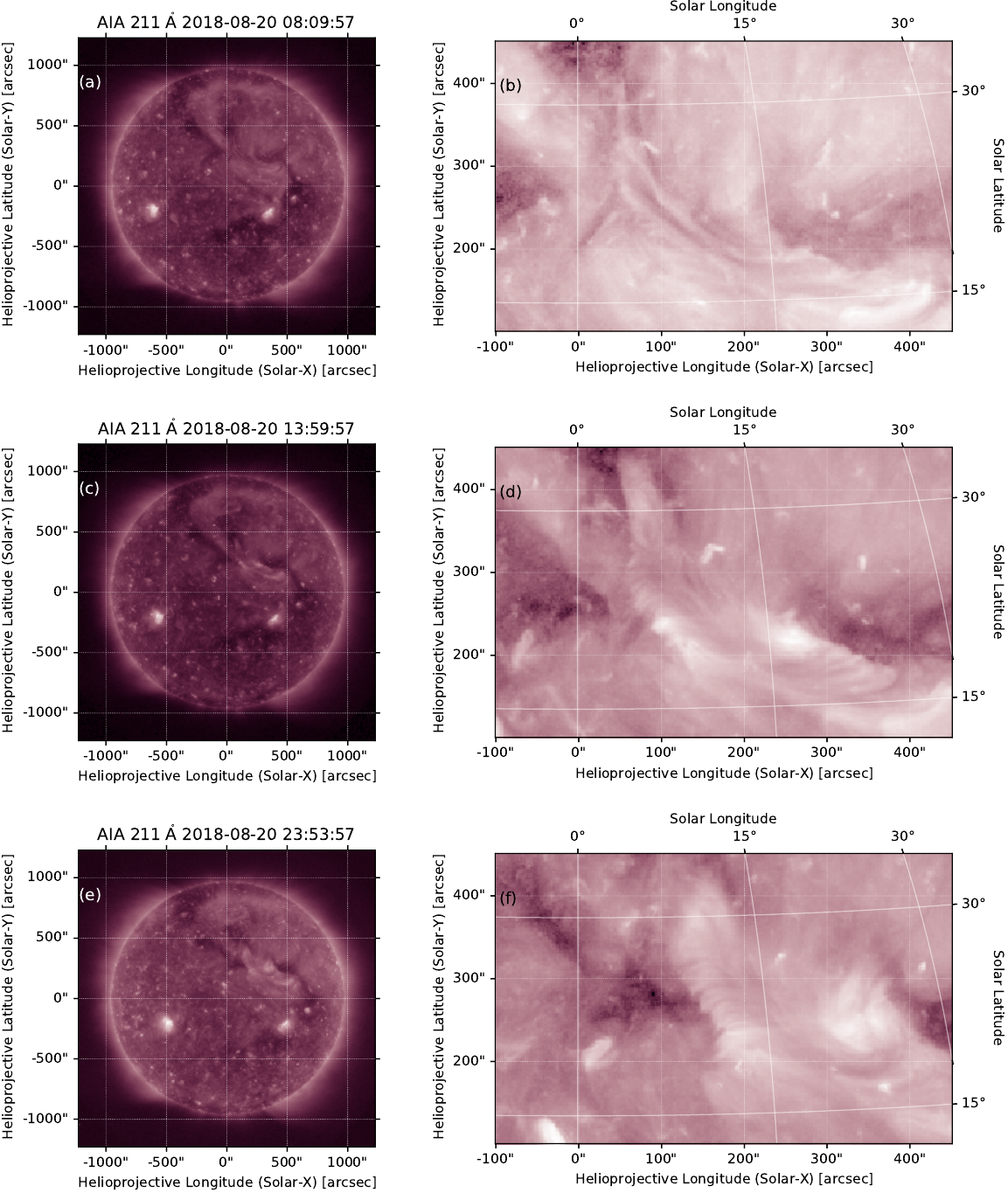}
 \caption{The SDO/AIA 211 \AA{}  full disc images (a, c, d) and the zoomed view (b, c, d) during 
different stages of the filament eruption that occurred on 20 August 2018.}
\end{figure}
\end{landscape}

\begin{figure*}
\centering
 \includegraphics[width=1.5\columnwidth]{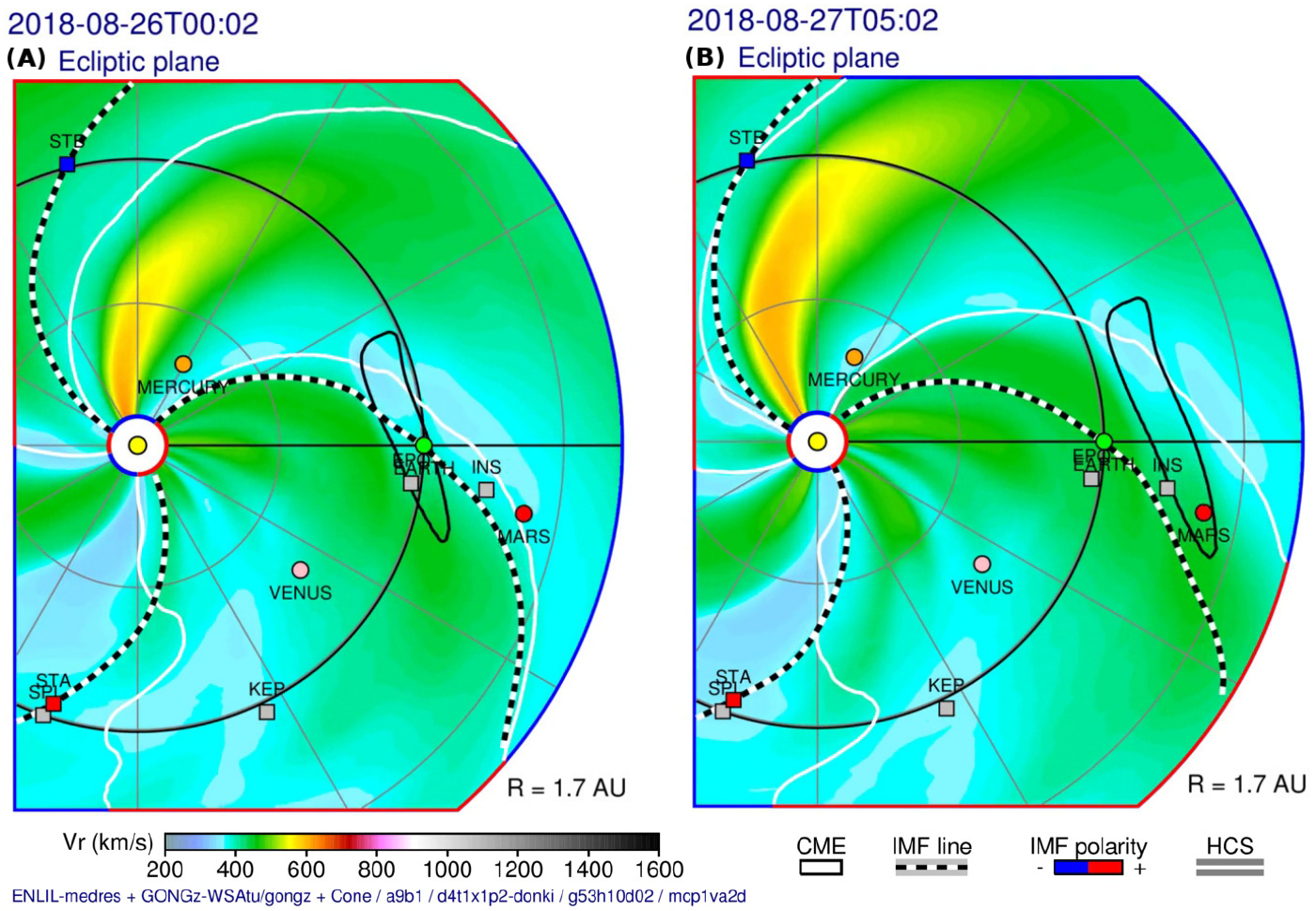}
  \caption{The WSA-ENLIL+Cone model inner heliospheric simulation snapshots showing the solar wind 
radial velocity (color contour) and IMF during stealth CME event of August 2018. The relative 
positions of Earth and Mars are also shown.}
\end{figure*}

\begin{figure*}
\centering
 \includegraphics[width=2\columnwidth]{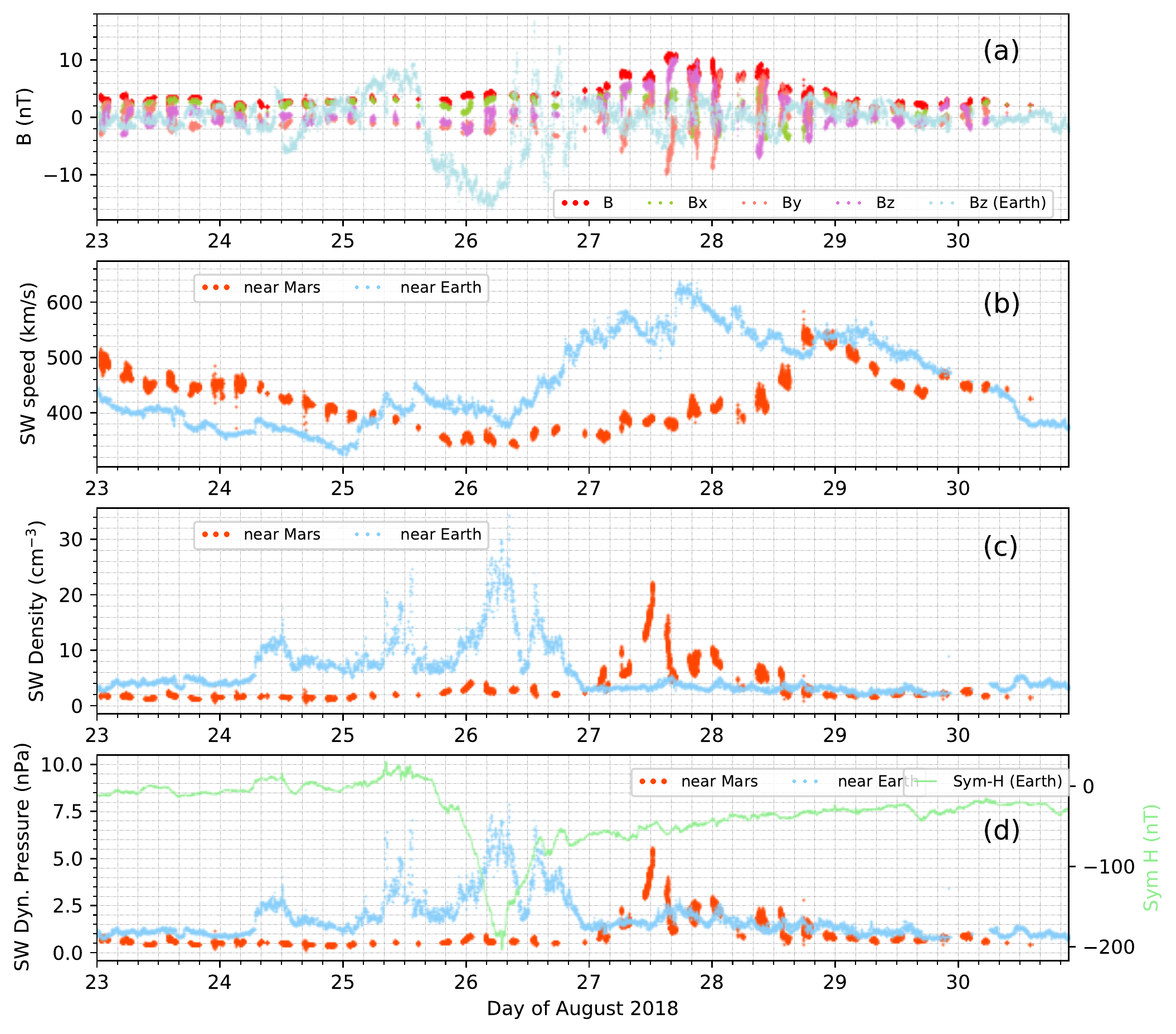}
  \caption{IMF (a), solar wind speed (b), solar wind density (c) and dynamic pressure (d) 
observations during 23-31 August 2018, near Earth and Mars. The Sym-H variation, indicating the 
occurrence of an intense geomagnetic storm at Earth is also shown in (d).}
\end{figure*}

\begin{table*}
\centering
\begin{tabular}{|c |c |c |c |c | c| c| c| c| c| c| c| c|}
\hline 
{Day/} & {UTC} & {UTC} & {Alt} & {Alt} & {Lon} & {Lon} & 
{Lat} & {Lat} & {SZA} & {SZA} & {LST} & {LST} \\ 
{Orbit} & {(hr)} & {(hr)} & {(km)} & {(km)} & {(deg)} & 
{(deg)} & 
{(deg)} & {(deg)} & {(deg)} & {(deg)} & {(hr)} & {(hr)} \\
\hline 
{INBOUND} & From & To & From & To & From & To & From & To & From & To & From & To\\
\hline

27/Orbit 1 & 0.25   &   0.45  &  499.10  &  149.45   &  19.68  &  316.06  &   74.85   &  41.15 &   116.61 &   160.37  &   19.58   &   0.02 \\

27/Orbit 2 & 4.67   &   4.87  &  496.45  &  150.69  &   84.17  &   20.91  &   74.82  &   41.31  &  116.66  &  160.24  &   19.59  &   23.99 \\

27/Orbit 3 & 9.10  &    9.29  &  497.58  &  150.37  &  149.94 &    85.68  &   74.87  &   41.31  &  116.36  &  160.26  &   19.50  &   23.98 \\

27/Orbit 4 & 13.52  &   13.71 &   498.14  &  150.14 &   215.40 &   150.50 &    74.92  &   41.51  &  116.13  &  160.09   &  19.44  &   23.96 \\

27/Orbit 5 & 17.94  &   18.13  &  497.02  &  150.64 &   280.43 &   215.33  &   74.89  &   41.55  &  116.04  &  160.07  &   19.41  &   23.94 \\

27/Orbit 6 & 22.36  &   22.55 &   497.73  &  149.97 &   346.11 &   280.18  &   74.95  &   41.79 &   115.75 &   159.85  &   19.33  &   23.91 \\

28/Orbit 1 & 2.78   &   2.97  &  496.78  &  150.48  &   51.06  &  344.97  &   74.94 &    41.83 &   115.67  &  159.82   &  19.30  &   23.90 \\

28/Orbit 2 & 7.20  &    7.39  &  499.24  &  151.24  &  116.86  &   49.87  &   74.97  &   42.12  &  115.36 &   159.53  &   19.22  &   23.87 \\

28/Orbit 3 & 11.62  &   11.82 &   498.03  &  150.43  &  182.27 &   114.69  &   75.00 &    42.26  &  115.15  &  159.40  &   19.16  &   23.85 \\

\hline

24/Orbit 1 & 1.51   &   1.71  &  498.63  &  148.72  &   54.51  &  358.80  &   74.24 &    38.79  &  119.58 &   161.72  &   20.44   &   0.35 \\

24/Orbit 3 & 10.35  &   10.55 &   499.85 &   148.37  &  185.58 &   128.40  &   74.39  &   38.97  &  119.07 &   161.71  &   20.31   &   0.31 \\

24/Orbit 4 & 14.78  &   14.97  &  499.64  &  149.36 &   250.48  &  193.31  &   74.37  &   39.25  &  119.02 &   161.53  &   20.28  &    0.28 \\

24/Orbit 5 & 19.20  &   19.39 &   496.73 &   148.92  &  315.40  &  258.08  &   74.40  &   39.26 &   118.94 &   161.60  &   20.26  &    0.27 \\

25/Orbit 1 & 4.04   &   4.24  &  497.45 &   149.76  &   85.92  &   27.73  &   74.47  &   39.54  &  118.61  &  161.47  &   20.16  &    0.23 \\

25/Orbit 2 & 8.46   &   8.66  &  498.66  &  149.28 &   151.70  &   92.61  &   74.54  &   39.82  &  118.30   & 161.27  &   20.07   &   0.20 \\

25/Orbit 3 & 12.88  &   13.08  &  496.55 &   149.33 &   216.44 &   157.34  &   74.56  &   39.80 &   118.26  &  161.34   &  20.06  &    0.19 \\

25/Orbit 4 & 17.30  &   17.50 &   498.70 &   149.54 &   282.32 &   222.28  &   74.60  &   40.10  &  117.94  &  161.11  &   19.97   &   0.16 \\

25/Orbit 5 & 21.73  &   21.92 &   498.35  &  148.81 &   347.69 &   287.08  &   74.67  &   40.26  &  117.73  &  161.01  &   19.91   &   0.14 \\

26/Orbit 2 & 6.57  &    6.76 &   496.86  &  150.10 &   117.76  &   56.79  &   74.68  &   40.64 &   117.53  &  160.73  &   19.84  &    0.10 \\

26/Orbit 4 & 15.41  &   15.60  &  496.48  &  150.32 &   248.17 &   186.42  &   74.73  &   40.87  &  117.22  &  160.59   &  19.75  &    0.06 \\

26/Orbit 5 & 19.83  &   20.03  &  497.04  &  149.98  &  313.89 &   251.19  &   74.79  &   40.86  &  116.93  &  160.63   &  19.67   &   0.04 \\

\hline
\end{tabular}
\label{table1}
\bigskip
\caption{Periapsis pass time in UTC (with day of August 2018 and orbit of the day), altitudes, 
longitudes, latitudes, SZA, and LST for disturbed orbits (27/28 August) and representative quiet 
orbits (24, 25, and 26 August) during the inbound legs of MAVEN [measurement below 500 km 
altitude].}
\end{table*}

\begin{figure*}
\centering
 \includegraphics[width=1.1\columnwidth]{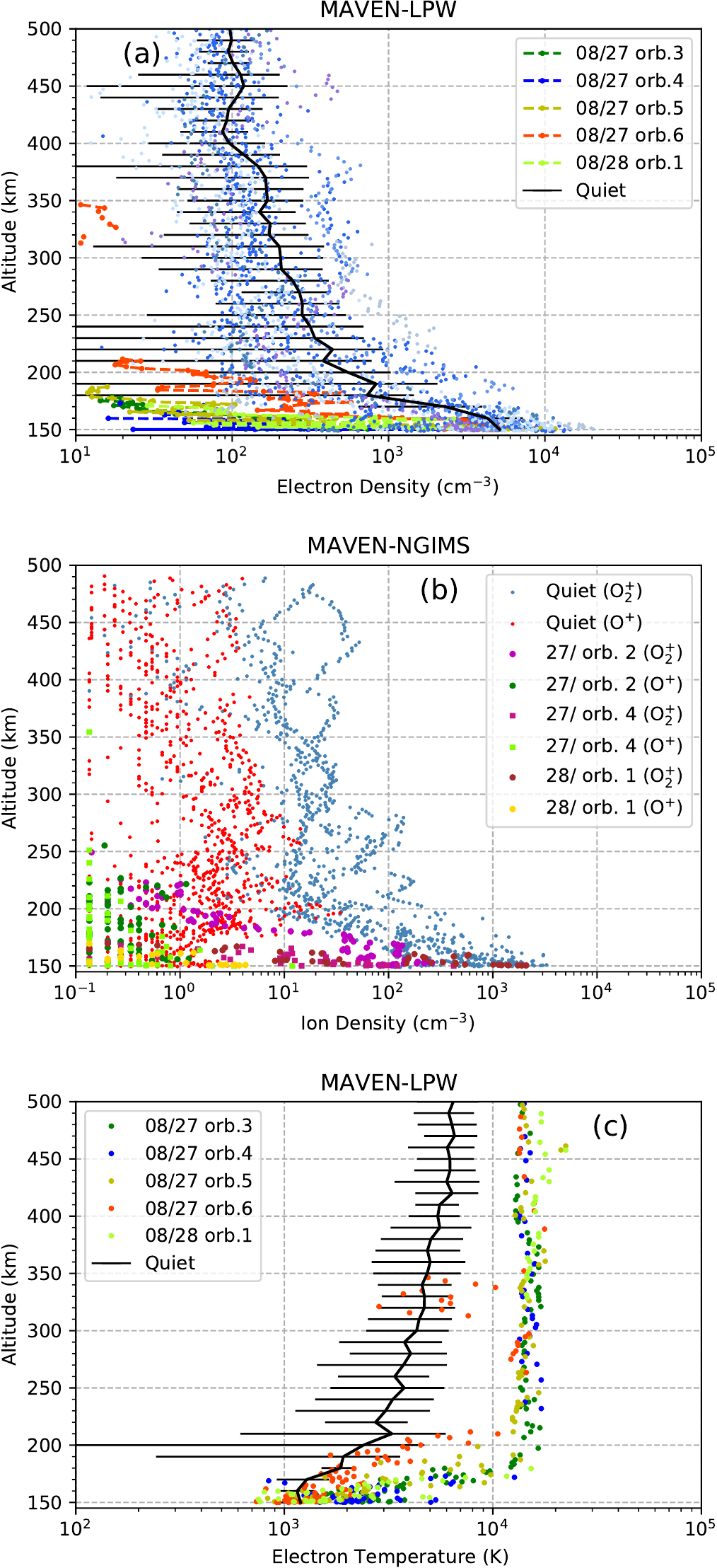}
  \caption{(a) The LPW  observations during 27-28 August 2018, along with the typical quiet time 
variation. The 7 quiet orbits on 24, 25 and 26 August 2018, are shown as blue dots. The mean of the 
quiet time profiles is shown (black line) along with standard deviation. (b) The NGIMS O$^+$ (amu 
16), and O$_2^+$ (amu 32) observations during 27-28 August 2018, along with the typical quiet time 
variation. The quiet time variations are from observations on 26 August 2018. (c) The LPW  Electron 
temperature estimates during 27-28 August 2018, along with the quiet time variation. Both LPW and 
NGIMS observations  are during the the inbound phase of the MAVEN spacecraft.} 
\end{figure*}





\end{document}